\documentstyle[preprint,prl,aps,epsf]{revtex}  

\begin{document}                
\title{First experimental evidence of 
one-dimensional \\ plasma modes in superconducting thin wires}
\author{ 
B. Camarota$^a$, 
F. Parage$^a$, F. Balestro$^a$, P. Delsing$^b$ and \\ O. Buisson$^a$}
\address{
$^a$ Centre de Recherches sur les Tr\`es Basses Temp\'eratures, 
Laboratoire Associ\'e \`a l'Universit\'e\\Joseph Fourier, 
C.N.R.S., BP 166, 38042 Grenoble-C\'edex 9, France.\\ 
$^b$ Department of Microelectronics and Nanoscience, Chalmers 
University of Technology and Goteborg University, S-412 96, Goteborg, 
Sweden.}

\maketitle
\begin{abstract}
We have studied niobium superconducting thin wires deposited onto a 
SrTiO$_{3}$ substrate. By measuring the reflection coefficient of the 
wires, resonances are observed in the superconducting state in the 
130~MHz to 4~GHz range. They 
are interpreted as standing wave resonances of one-dimensional
plasma modes propagating along the superconducting wire. 
The experimental dispersion law, $\omega$ versus $q$, presents a linear 
dependence over the entire wave vector range. The modes are softened as the temperature increases 
close the superconducting transition temperature. Very good agreement 
are observed between our data and the dispersion relation predicted by 
Kulik\cite{Kulik74} and Mooij and Sch\"on\cite{Mooij85} .
\end{abstract}


\vskip 1.0truecm
\narrowtext


Quantum phenomena in mesoscopic physics such as tunnelling effects, 
quantum fluctuations or decoherence processes are intimately related to the description of the 
environment. Its effects on quantum systems were
extensively discussed in the last two decades. Dissipative 
environment was taken into account in the seminal Caldeira-Leggett 
study in order to treat the macroscopic quantum tunneling in a 
titled washboard potential\cite{Caldeira81}. Slightly afterwards quantum 
phase transition was predicted for Josephson junctions in the presence of a resistive 
environment~\cite{Schmid83}.

In most of Josephson junctions experiments, the dissipative environment corresponds 
to external circuits made by the measurement circuit and is therefore extrinsic to the quantum 
system.
In contrast in thin superconducting wires, the environment is intrinsic to the
system. The related modes are the one-dimensional (1D) plasma 
modes propagating along the wire. They were predicted for the first time in a superconducting 
wire by 
Kulik\cite{Kulik74} and 
later analysed by Mooij and 
Sch\"on using non-equilibrium superconductivity model~\cite{Mooij85}. The restoring force is 
the long range Coulomb interaction. Because of the restricted geometry, the 
charge mode 
is not shifted to the bulk plasma frequency and has a sound like 
dispersion relation.
Due to the gapless dispersion law of the environmental modes,
quantum fluctuations in superconducting thin wires show critical 
behaviours \cite{Giordano88,Duan95,Bezryadin00}. A 
new quantum superconducting insulator phase transition has recently been 
predicted\cite{Zaikin97} and observed in ultra-thin 
nano-wires\cite{Bezryadin00}. The driving parameter of the phase transition is the 
friction term coming from 
the interaction between the quantum system and the environmental modes.

In a thin superconducting loop closed by a Josephson junction, 
quantum fluctuations of the propagating plasma modes are 
predicted to renormalize the Josephson energy of the junction 
\cite{Hekking97}. Furthermore these collective excitations are 
expected to modify IV characteristics of normal metal-superconductor 
tunnel junction \cite{Kim93,Falci95}.
In Josephson junction arrays the plasma modes, also called 
spin-waves, can affect the quantum dynamics of 
vortices~\cite{Eckern89,Vanderzant93,Fazio94}.

Although many theoretical and experimental works have pointed out the importance 
of one-dimensional~(1D) plasma modes in thin superconducting wires as 
environmental modes, no experiments 
have succeeded in observing them yet.
Indeed up to now, to our knowledge, only two-dimensional plasma modes have 
been measured 
in superconducting granular aluminium films~\cite{Buisson94}, later in 
YBCO films~\cite{Dunmore95} and more recently in superconducting wires 
networks~\cite{Parage98}.

In this letter we report the first experimental evidence of 
propagating 1D-plasma 
modes in superconducting wires. The wires were deposited onto 
strontium titanate (SrTiO$_3$) crystal.
The  configuration "superconducting film/SrTiO$_3$" was proven to be a
{\it model experimental system}  to study general properties of plasma modes 
\cite{Buisson94,Parage98}: in fact superconducting  properties give a 
very weak damping of plasma oscillations, moreover the high dielectric 
constant of the SrTiO$_3$ $(\epsilon_m\approx$ 10$^{4}$ at low 
temperatures) reduces their energy by two orders of magnitude.

Three different wires, $A$, $B$ and $C$ were measured.
Each one was deposited onto a (110) SrTiO$_3$ 
substrate of thickness H~=~0.3~mm. 
The wires were obtained from very thin niobium films of about 10~nm 
thickness 
evaporated in a ultrahigh vacuum chamber at room temperature.
A 5~nm-thin layer of silicon was added in order to protect the 
niobium during the lithography process.
The wire pattern was defined on a 130-200~nm thick negative Nover 
\cite{Nover} resist layer by electron-beam lithography.
Both silicon and niobium layers were etched in a 
SF$_{6}$ plasma. Next the Nover resist was removed by oxygen plasma leaving 
the niobium pattern. 
The dimension of the wires such as their width, $W$, 
thickness, $t$, and total length, $L$, are given in Table I.
In order to perform electrical bonding, the wire is widened
to a 30~$\mu$m width over a 100~$\mu$m length at its two extremities.

Table I summarizes the main geometrical parameters and physical 
properties of the three different measured wires. The critical temperature
of the superconducting transition was 
obtained from transport measurements and taken at the midpoint of the 
resistive transition. The film resistivity $\rho$ was measured just 
above the superconducting transition. The penetration depth 
$\lambda_{\rm th}(0)$ was derived 
from the BCS dirty model, $\lambda_{\rm th}(0)=\lambda_{\rm 
L}(0)(\xi'_{0}/l)^{1/2}$ where $\lambda_{\rm 
L}(0)$~=~37~nm is the London length, $\xi_{0}'=62~nm (9.25~K/T_{c})$ 
the modified Pippard coherence length~\cite{Halbritter74} and $l$ the mean free 
path deduced from the resistivity by taking for niobium $\rho 
l=0.72\,\,10^{-5}\mu\Omega.cm^2$.

In order to observe plasma modes resonances, reflection coefficient 
measurements were performed in the 130~MHz to 4~GHz range using an HP8720B vector analyzer.
A cryogenic 50$\Omega$ coaxial 
cable guides the microwave between the vector analyser and the 
sample. The excitation of plasma modes is realized by injecting 
external charges  into one extremity of the
superconducting thin wire at the  
microwave frequency. The electrical contact between the wire and the inner conductor of the coaxial line 
is made by means of a 20$\mu m$-diameter aluminum bonding wire. The 
"superconducting wire/SrTiO$_3$" block is isolated from electrical 
ground plane by a thin teflon slab.

 Since similar results have been obtained on the three different samples, 
only the sample $A$ will be reported in detail. Typical reflection coefficient versus frequency 
is plotted in Fig.~1 between 130~MHz and 2~GHz.
Resonances show two distinct behaviours.
Above  about 1.5~GHz, large resonant peaks are observed. They are 
weakly temperature dependent and exist both in the normal and 
the superconducting state.
These resonances are related to dielectric modes inside the 
SrTiO$_3$. They appear above the cut frequency of the 
Transverse-Electric mode  which is estimated at about
1.5~GHz for the configuration used in this experiment.
Much more interesting are the resonances which appear below 1.5~GHz.
Their amplitude and frequency are strongly temperature dependent. As 
the temperature increases, their peaks shift towards lower frequency, 
their amplitude decreases and 
their width increases. In the normal state, these resonances disappear.

Only two possible modes could explain the resonances below 1.5~GHz
: the Transverse Electro-Magnetic (TEM) modes 
and the 1D plasma modes.
Confusion between these two modes could be made because 
they both verify similar properties such as a quasi-linear 
dispersion law and a temperature dependence related to the superfluid 
density. However fundamental differences exist between them. 
Plasma modes need only one wire to propagate while TEM modes need the 
wire and a ground plane. The 
quasi-linear dispersion of 1D plasma modes comes from the long 
range Coulomb interaction of the wire whereas the linear 
dispersion of the TEM modes is explained by the short range Coulomb 
interaction of the wire screened by the ground plane. The following analysis of 
the experimental configuration can explain why TEM modes can not be 
taken into account to interpret our experimental results.

The superconducting wire/SrTiO$_3$/teflon slab/ground plane structure 
\cite{CamJLTP2000} corresponds to a
microstrip whose the inner conductor is the wire and is separated from the ground plane by
SrTiO$_{3}$  and teflon slab dielectrics.
The TEM mode related to this structure is determined by the unit-length
inductance, $L_l$, and capacitance, $C_l$~\cite{Microstrip}. The capacitance corresponds 
to the capacitance between the wire and the ground plane through the 
two dielectrics. By measuring the impedance of the wire at 130~MHz 
using the network analyser, the capacitance could be deduced, 
$C_l\approx$0.45~nF/m. 
Since the  wire is superconducting, the inductance $L_l$ is the sum of two terms: 
one is related to the kinetic inductance, estimated at about L$_{K}\approx$5.2$\mu$H/m for 
a penetration depth of about 0.5$\mu$m; the other is related to the magnetic inductance of the 
microstrip 
(L$_{m}\approx$1.3$\mu$H/m) \cite{Microstrip}. For such parameters, 
the phase velocity of the TEM modes, given by $1/\sqrt{L_lC_l}$, is 
roughly 2 10$^{7}$ m/s.
Thus, the first 
resonance related to the TEM mode propagating along the wire
is expected at about 4~GHz, one order of magnitude higher than 
the observed ones. Therefore standing wave resonances of TEM modes 
can not explain the 
resonances measured below 1.5~GHz.

 Hence we will discuss the 
observed resonances as standing wave resonances of 1D plasma modes.
Indeed as it was already shown in a previous study \cite{Buisson94}, the wave 
vectors associated to plasma resonances obey the $\,q\,L=\,n\pi$ selection rule for standing 
waves, where $n$ an integer indexing the different resonances.
An experimental $\omega$ vs. $q$ plasma 
dispersion relation is therefore obtained. Fig. 2 presents the dispersion law 
of sample A for different temperatures. The linear 
dependence of the dispersion relation is observed over the entire wave 
vector range and for temperatures going from 1.5~K to very close to T$_{c}$.
As the temperature increases, the modes are strongly 
softened. The mode velocity varies from  2~10$^{5}$ m/s
near T$_{c}$ up to  1.5~10$^{6}$ m/s at low temperature, but is 
always slower than the light velocity in the SrTiO$_3$. Above T$_{c}$,
no dispersion curve exists because of the 
disappearance of the resonances.

Theoretical studies on plasma modes in superconducting 
wires~\cite{Kulik74,Mooij85}
predict a quasi-linear dispersion relation if 
the radius $r_{0}$ of the wire is much thinner than the penetration depth, 
$\lambda(T)$, and the wavelength ($r_0\ll\lambda(T)$ and 
$\,q\,r_0\ll1$).
 Plasma modes dispersion relation is then given by:

\begin{eqnarray}
\omega_p^2 =  {{r_0}^2\over{\epsilon_0\,\mu_0\,\epsilon_m
\,\lambda^2(T)}}\; \tilde{q}^2\ln(1/(\tilde{q} r_0))   
\label{disp1D}
\end{eqnarray}
where $\tilde{q} = \sqrt{q^2- \mu_0\,\epsilon_0\,\epsilon_m 
\omega_{p}^2}$ is the parameter characterizing evanescent length inside 
the dielectric, taking into account retardation 
effects. Such theoretical results were obtained for a cylindrical shaped wire 
embedded in an infinite isotropic medium of dielectric constant 
$\epsilon_m$.

The experimental configuration presented here shows some important 
differences with the one considered in the theoretical derivations. Indeed 
the SrTiO$_3$ is finite size and a ground plane is present close to 
the superconducting wire. 
One-dimensional plasma modes are very sensitive to the finite 
dimension of the sample 
because of the long range Coulomb interaction which extends up to few 
millimeters if very low energy plasma modes are considered. Their propagation 
may 
thus be expected to be strongly modified.

In order to take into account these perturbations we have derived, in a previous work~\cite{CamJLTP2000},
the plasma modes dispersion 
relation of a wire in the real experimental 
configuration.
Two simplifications were applied. Firstly the SrTiO$_3$ 
dielectric constant was assumed isotropic with an average value
$\epsilon_m$~= $(\epsilon_{001}\epsilon_{110}\epsilon_{1\bar{1}0})^{1/3}$.
Secondly the rectangular wire was 
approximated by a cylindric wire of radius $r_0$ with cross section 
$\pi r_0^2/2$=$\,W\,t$ and half-immersed in the semi-space occupied by the 
SrTiO$_3$.
Because of the very high dielectric constant of SrTiO$_3$ 
($\epsilon_m\approx$12500 for sample A), it was possible to take into 
account the finite size of the experimental configuration. 
As a result the derived dispersion 
relation of our sample configuration~\cite{CamJLTP2000} was proven to be well described by the predicted 1D 
plasma modes one, Eq. 1, as long as the evanescent length 
is smaller than twice the SrTiO$_{3}$ thickness ($2\tilde{q}H>1$).
This allowed us to conclude that the experimental set-up represents an optimal 
configuration for the study of plasma modes in superconducting wires.

The comparaison between the experimental dispersion law and 
the theoretical one given by Eq.~\ref{disp1D} is shown in Fig.~2. The 
theoretical quasi-linear 1D plasma mode dispersion fits very well the experimental 
one for all the temperatures and over the entire wave vector range. 
In the range of 
wave vectors accessible by this experiment, the logarithmic deviation 
predicted by the theory is not observed.
Since $\epsilon_m$ is measured independently using the dielectric 
resonances~\cite{Parage97}, $\lambda(T)$ is the only free parameter.
The experimental 
penetration depth extracted from the fit versus the reduced 
temperature T/T$_{c}$ is plotted in Fig.~3 for the 
three different samples. We notice that the required condition 
$2\tilde{q}H>1$ is fullfilled for sample A and B over all the 
temperature range but not for sample C below 8~K justifying that the penetration 
depth of sample C is plotted only near T$_{c}$. 
As the temperature increases close 
to T$_{c}$, the penetration depth diverges. At low temperature 
(T/T$_{c}<0.5$), $\lambda(T)$ saturates. The temperature dependence 
cannot be well fitted by the Gorter-Casimir law but by an equivalent 
one: 
$\lambda(T)=\lambda_{exp}(0)/\sqrt{1-(T/T_{c})^2}$ where $\lambda_{exp}(0)$ is the 
only free parameter. The same temperature 
dependence has also been found in previous studies on niobium wire 
networks~\cite{Parage98,Parage97}. The penetration depths
at T=0~K obtained from the fit, $\lambda_{exp}(0)$, are summarized in Table I for the three 
samples. These experimental values are consistent with the 
penetration depth derived using the BCS dirty limit.

In conclusion one-dimensional plasma modes have been observed for the 
first time by measuring the reflection coefficient of thin niobium 
wires. By using a very high 
dielectric constant substrate, the stontium titanate, the long range Coulomb interaction is strongly 
weakened, reducing by two orders of magnitude the plasma modes energy.
A dispersion law has been obtained which shows a linear dependence. 
Moreover the plasma modes are softened as the temperature increases 
close to the superconducting transition temperature. Both the dispersion law and the 
temperature dependence are well explained by the predicted plasma 
modes dispersion given by Eq.~1.

\section*{Acknowledgments}
{We thank M. Doria, G. Falci and F.W.J Hekking for invaluable discussions.
This research is supported by european Union under TMR 
Program "Dynamics of nanofabricated superconducting circuits".

\begin{center}
\begin{table}
\caption{Characteristics of the different niobium 
wires.\hspace{1cm}}\label{tab1}
\begin{tabular}{|c|c|c|c|c|c|c|c|c|c|}
Samples & $t$ (nm) & $W$ ($\mu$m) & $L$ (mm) & $T_c$ (K) & $\rho$ 
($\mu\Omega$.cm) & $l$ (nm) & $\xi'_{0}$ (nm) & $\lambda_{\rm th}(0)$ ($\mu$m) & $\lambda_{\rm exp}(0)$ ($\mu$m) \\ \hline 
 A & 12 & 5.0 & 2.2 & 5.19 & 55 & 1.3 & 110 & 0.34 & 0.50 \\
 B & 11 & 3.0 & 2.25 & 6.68 & 32.4 & 2.2 & 86 & 0.23 & 0.21\\
 C & 10 & 6.0 & 2.38 & 8.84 & 16.8 & 4.3 & 65 & 0.14 & 0.14\\ 
\end{tabular}
\end{table}
\end{center}

\begin{figure}
\centerline{\epsfxsize=10 cm \epsfbox{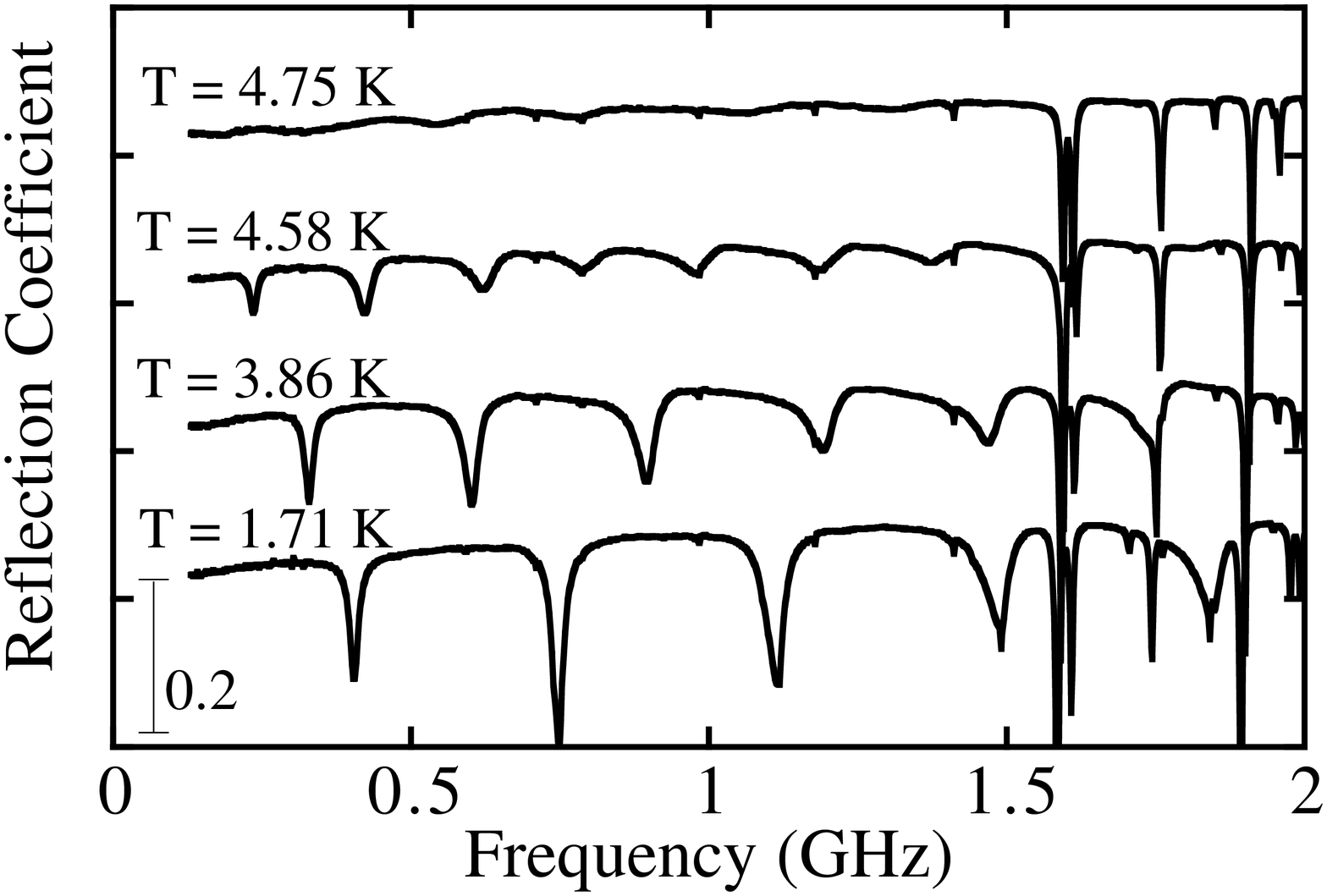}}
\baselineskip 10pt {\small{\bf Fig.1.} Reflection coefficient versus 
frequency at different temperatures.
A vertical shift of the different curves have been introduced for clarity. }\\
\label{reflcoeff}
\end{figure}

\begin{figure}
\centerline{\epsfxsize=10 cm \epsfbox{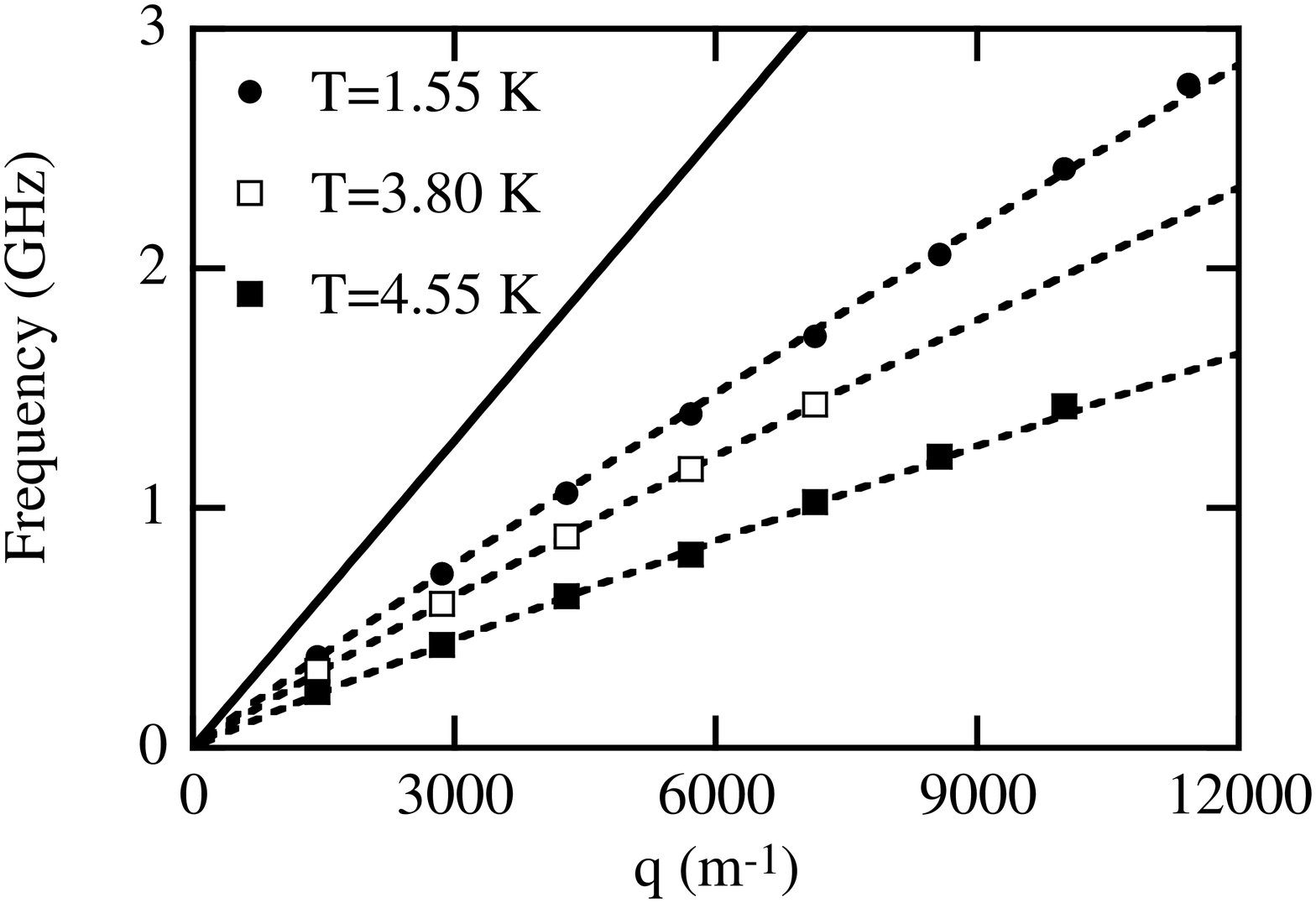}}
\baselineskip 10pt {\small{\bf Fig.2.} Dispersion 
relation of sample A measured at three different temperatures (points) and fitted 
by the theoretical dispersion law given by Eq. 1 (dashed lines). Light 
dispersion inside the SrTiO$_{3}$ (continuous line) is plotted for comparaison.}
\label{dispersion1D}
\end{figure}



\begin{figure}
\centerline{\epsfxsize=10 cm \epsfbox{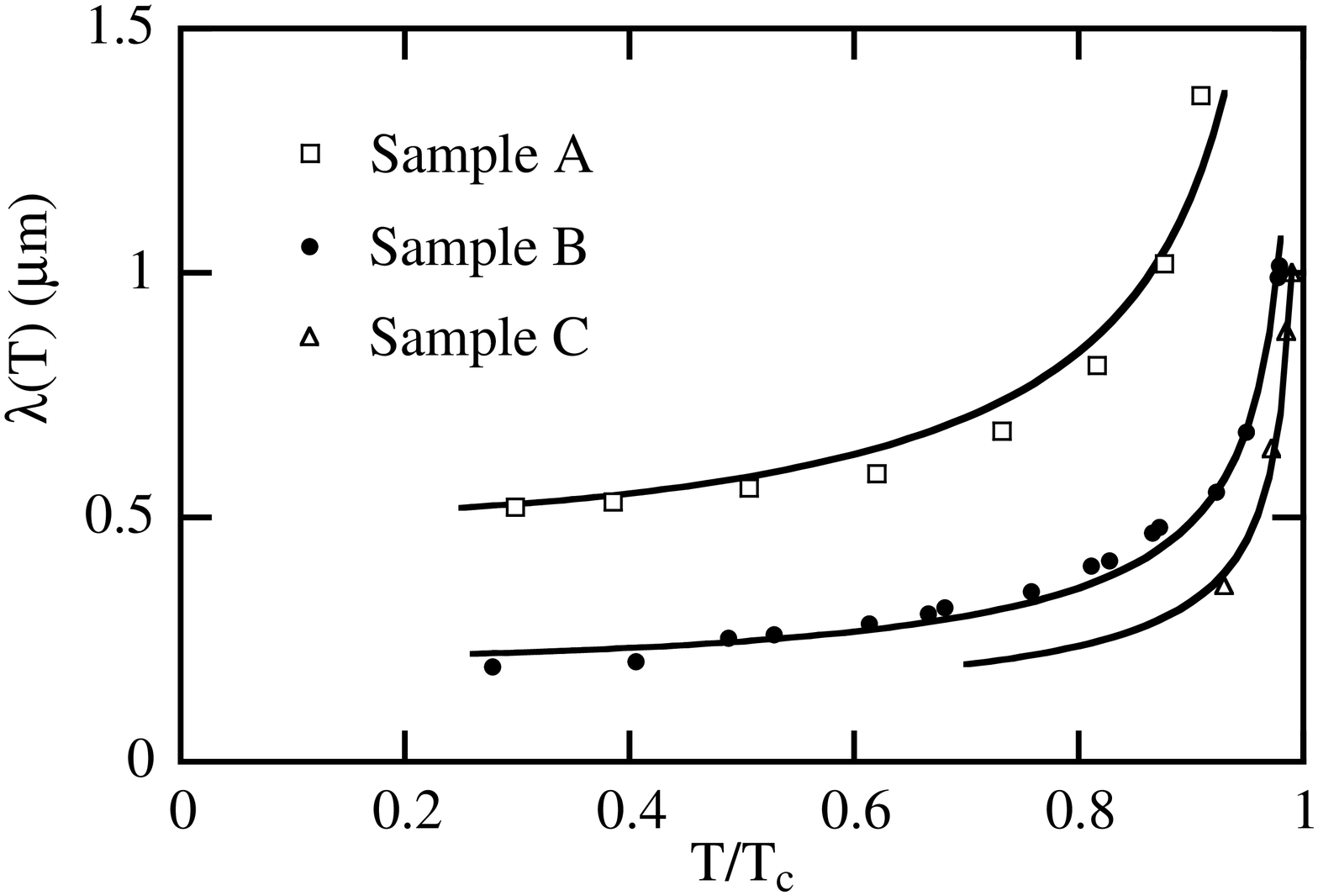}}
\baselineskip 10pt {\small{\bf FIG. 3.} Experimental penetration depth (dots) deduced from the fit of Eq.1 
as a function of the reduced temperature T/T$_{c}$. Solid lines are obtained from 
$\lambda(T)=\lambda_{exp}(0)/\sqrt{1-(T/T_{c})^2}$ with  
$\lambda_{exp}(0)$ as adjustable parameter (see values in Table I).}\\
\end{figure}

\widetext

\end{document}